\documentclass[aps,floatfix,prb,10pt,twocolumn,showpacs,superscriptaddress,notitlepage]{revtex4-1}
\usepackage[utf8x]{inputenc}
\usepackage{graphicx}
\usepackage{amsmath}
\usepackage{lipsum}
\usepackage{hyperref}
\usepackage{placeins}
\graphicspath{{./figure/}}

\begin{document}

\title{Inherent Structures in \texorpdfstring{\boldmath$m$}{m}-component Spin Glasses}

\author{M. Baity-Jesi} \affiliation{Departamento
  de F\'\i{}sica Te\'orica I, Universidad
  Complutense, 28040 Madrid, Spain.} \affiliation{Dipartimento di Fisica, La Sapienza Universit\`a di Roma,
     00185 Roma,  Italy.}\affiliation{Instituto de Biocomputaci\'on y
  F\'{\i}sica de Sistemas Complejos (BIFI), 50009 Zaragoza, Spain.}
\author{G.~Parisi}\affiliation{Dipartimento di Fisica, IPCF-CNR, UOS
Roma Kerberos and INFN, La Sapienza Universit\`a di Roma, 00185 Roma,  Italy.}

\date{\today}

\begin{abstract}
We observe numerically the properties of the infinite-temperature inherent structures 
of $m$-component vector spin glasses in three dimensions.
An increase of $m$ implies a decrease of the amount of minima of the free
energy, down to the trivial presence of a unique minimum. 
For little $m$ correlations are small and the dynamics are quickly arrested, 
while for larger $m$ low-temperature correlations crop up and the convergence is slower,
to a limit that appears to be related with the system size.
\end{abstract}

\maketitle

\section{Introduction}
Although it is established that typical spin glasses \cite{Mezard:87} order at a critical 
temperature $T_\mathrm{SG}$ for $d\ge 3$ \cite{Ballesteros:00,Kawamura:01,Lee:03},
the nature of the low-temperature phase of spin glasses under the upper critical dimension
$d_u=6$ is still a matter of debate \cite{Moore:11, Parisi:12, Yeo:12}. 
There are two main theories describing the spin glass phase. 

(1) The droplet theory portrays spin glasses as disguised ferromagnets,
with only two low-temperature states, and no spin glass phase once an external
magnetic field $h$ is applied \cite{Fisher:86, Bray:87, Fisher:88}.
Ferromagnets in disguise can be obtained, for example, by performing a random gauge 
transformation on an ordered system \cite{Nishimori:01}, as it is done in the Mattis model \cite{Mattis:76}.

(2) On the other side the replica symmetry breaking (RSB) theory 
characterizes the spin glass phase with a rugged free-energy landscape with 
an exponentially large number of states with an ultrametric structure \cite{Parisi:79,Parisi:80, Parisi:83}. 
Once the temperature is lowered
under the critical point, the phase space is split up in many regions 
separated by barriers that diverge with the system size, so ergodicity is broken.
When a field is applied, at high dimensions a critical 
line $h_c(T)$ (the dAT line \cite{Almeida:78}) separates the spin glass from the paramagnetic phase. 
In low enough dimensions both the transition at $h=0$ and $h\ne 0$ disappear. 

At the present moment the lower critical dimension at $h=0$ seems to be around $d=2.5$. 
Its value in a field, instead, is still matter of debate: The situation could be similar to that 
of a ferromagnet where the lower critical dimension in absence of a field (i.e. 1) is different 
from the lower critical dimension in the presence of a random field (i.e. 2).

Since increasing the number of spin components $m$ reduces the number of metastable states,
recent works focused on the properties of these models in the $m\rightarrow\infty$ limit, and their
energy landscape \cite{Hastings:00}. Interesting features have been pointed out in large-$m$ mean
field models, such as a Bose-Einstein condensation in which the spins condense from an $m$-dimensional 
to an $n_0$-dimensional subspace, with $n_0\sim N^{2/5}$ ~\cite{Aspelmeier:04}.

It has been argued in Ref. \onlinecite{Aspelmeier:04} that the $m=\infty$ limit could be a good
starting point for the study of the low-$m$ spin glasses \cite{Beyer:12}, via $1/m$ expansions that have been used,
for example, to try to question the presence of a dAT line \cite{Moore:12}. 
However the Hamiltonian of the $m=\infty$ model has a unique local minimum, that can be found easily 
by steepest descendent (the determination of the ground state is {\em not} an NP complete problem). 

Explicit computations  also indicate that the $m=\infty$ model is substantially different from
any finite-$m$ model (for example there is only quasi long-range order under $T_\mathrm{SG}$, the upper critical
dimension has been shown to be $d_u=8$, and the lower critical dimension is suspected to be $d_l=8$ too
\cite{Green:82,Viana:88,Lee:05}), and that it is more interesting to study these
models for large but finite $m$, thus reversing the order of the limits $m\rightarrow\infty$ 
and  $V\rightarrow\infty$ ($V$ is the total number of spins) \cite{Lee:05}.

To better understand the large (but finite) $m$ limit we undertake a numerical study in a three-dimensional 
cubic lattice. Our aim is to arrive at a quantitative comprehension of the energy landscape of systems with varying $m$, expecting, for example,
to observe growing correlations as $m$ increases \cite{Hastings:00}.

We focus on infinite-temperature inherent structures, i.e. the local energy minima that one reaches by
relaxing the system from an infinite-temperature state, that is equivalent to a random configuration. 
Examining a system from the point of view of the
inherent structures is a very common practice in the study of structural glasses \cite{Cavagna:09}.
Only recently the study of quenches
\footnote{By quench we mean the minimization of the energy throughout the best possible satisfaction
of the local constraints, i.e. a \emph{quench} is a dynamical procedure. 
In Sec.~\ref{sec:model-sim} we explain how the quenches were performed. Be careful
not to confuse it with other uses of the same term. For example, those quenches have little to do with
the \emph{quenched approximation} used in QCD, or the \emph{quenched disorder}, that is a property of the system.}
from a high to a lower temperature has stimulated interest 
also in spin systems, both in presence and absence of quenched disorder 
\footnote{Let us stress once again the meaning of the term quench, since the terminology might generate some confusion. 
We refer to \emph{quenched disorder} 
when the system is described by some random variables that do not change in time. This concept is independent from 
the \emph{quenches} we impose to our system, i.e. relaxing the system to the closest
local minimum of the energy.} \cite{Quenches,Berthier:04,Marco:11}.

We analyze the properties of the inherent structures,
and we inspect the dynamics of how the system converges to those configurations.

When one performs a quench from $T=\infty$ to $0<T=T_0<T_{SG}$, the system is expected to show
two types of dynamics, an initial regime where thermal fluctuations are irrelevant, and
a later one where they dominate the evolution (see for example the quenches performed in Ref. \onlinecite{Berthier:04}).
We choose $T_0=0$, so we can to show that the origin of the second dynamical regime is
actually due to thermal effects. We study the quenches as a function of $m$. While on one side
in the Ising limit $m=1$ the dynamics is trivial, and correlations never become larger than a single
lattice spacing, on the other side an increasing $m$ yields a slower convergence, with the arising of low-temperature correlations 
that we can interpret as interactions between blocks of spins.

The paper is structured as follows. In Sec.~\ref{sec:model-sim} we define our model and describe
how simulations were done. In Sec.~\ref{sec:obs} we define the observables we measured, and in
Sec.~\ref{sec:results} we show our numerical results. First we analyze the properties of the inherent
structures as a function of $m$, and then we show the time evolution of the observables during
the quenches. Conclusions are given in Sec.~\ref{sec:conc}. In the Appendix we show a simple example of how the selfoverlap
can depend on $m$.

\section{Model and Simulations}
\label{sec:model-sim}
\subsection{Model}
The model is defined on a cubic lattice of side $L$ with periodic boundary conditions. Each of the $V=L^3$ vertices 
$\vec x$ of the lattice hosts an $m$-dimensional spin $\vec\sigma_{\vec x}=(\sigma_{\vec x,1},\ldots,\sigma_{\vec x,m})$, 
with the constraint $\vec\sigma_{\vec x}\cdot\vec\sigma_{\vec x}=1$.
Neighboring spins $\vec\sigma_{\vec x}$ and $\vec\sigma_{\vec y}$ 
are linked through a coupling constant $J_{\vec{x},\vec{y}}$.
The Hamiltonian is 
\begin{equation}
\label{eq:H}
 {\cal H} = - \sum_{<\vec x,\vec y>}^V J_{\vec{x},\vec{y}}~\vec\sigma_{\vec x}\cdot\vec\sigma_{\vec y}\,,
\end{equation}
where the brackets $<\ldots>$ indicate that the sum goes only over the nearest neighbors. The
spins are our dynamic variables, while the couplings $J_{\vec{x},\vec{y}}$ are Gaussian-distributed,
with $\overline{J_{\vec{x},\vec{y}}}=0$ and $\overline{J_{\vec{x},\vec{y}}^2}=1$.
We define different realizations of the couplings in a lattice with the term \emph{samples}.
Independent configurations of the spins that have the same couplings are called \emph{replicas} of
the same sample. This Hamiltonian is invariant under the simultaneous rotation or reflection of all the spins [that belongs to the $O(m)$ 
symmetry group], so the energy minimas may be found modulo a global rotation.

\subsection{Simulations}
We are interested in the inherent structures from infinite temperature, hence we need to pick
random starting configurations, and directly minimize the energy.
\footnote{The couplings and the initial configurations were chosen at random with a combination of congruential 
and Parisi-Rapuano random number generator.\cite{Parisi:85,Fernandez:09b}}

The algorithm we choose is a direct quench, that consists in aligning each spin to its local field $\vec{h}_{\vec x}$.
 \footnote{The local field is defined as  
 $\vec{h}_{\vec x} = \sum_{{\vec y}:||{\vec x}-{\vec y}||=1} J_{{\vec x},{\vec y}} \sigma_{\vec y}$.}
This choice was done because it allows us to compare inherent structures from systems with a different $m$
in a general way,
\footnote{For example, the Successive Over Relaxation used in Ref.~\onlinecite{Marco:11}
yields inherent structures with different properties, depending on the value of a parameter $\lambda$,
and the same $\lambda$ is not equivalent for two different values of $m$.}
because it is the most simple and widely used in literature, and because it leads 
to inherent structures that are conceptually mostly similar to those conceived in the field of supercooled liquids.

For each sample we simulated two replicas, in order to be able to compute overlaps.
We fixed the number of full sweeps of a lattice to $N_\mathrm{t}=10^5$, as it had already been
done in Ref.~\onlinecite{Berthier:04} with quenches to finite temperature. This amount
of steps was enough to guarantee the convergence to an inherent structure in all our simulations.
To ensure the convergence we required the last (logarithmically spaced) measurements to be equal 
within the error for each of the measured observables.

In Tab.~\ref{tab:param} we give the parameters of our simulations.
\begin{table}[b]
\begin{ruledtabular} 
 \begin{tabular}{ccccc}
  $L$ & $m$ & $N_\mathrm{s}$ & $N_\mathrm{t}$ & $N_\mathrm{m}$\\\hline\hline
  8   & 1  & 10000 & $10^5$ & 22\\
  8   & 2  & 10000 & $10^5$ & 22\\
  8   & 3  & 10000 & $10^5$ & 22\\
  8   & 4  & 5000  & $10^5$ & 22\\
  8   & 6  & 10000 & $10^5$ & 22\\
  8   & 8  & 10000 & $10^5$ & 22\\\hline
 16   & 1  & 1000  & $10^5$ & 22\\
 16   & 2  & 1000  & $10^5$ & 22\\
 16   & 3  & 1000  & $10^5$ & 22\\
 16   & 4  & 1000  & $10^5$ & 22\\
 16   & 8  & 1000  & $10^5$ & 22\\
 16   & 12 & 1000  & $10^5$ & 22\\
 16   & 16 & 1000  & $10^5$ & 22\\\hline
 64   & 3  & 160    & $10^5$ & 22\\
 \end{tabular}
\end{ruledtabular}
\caption{Parameters of our simulations. $N_\mathrm{s}$ is the number of simulated samples, 
$N_\mathrm{t}$ is the number of quench sweeps of the whole lattice, and $N_\mathrm{m}$ is
the number of measures we did during the quench. We chose to follow the same roughly logarithmic
progression chosen in Ref.~\onlinecite{Berthier:04}, measuring at times 2, 3, 5, 9, 16, 27, 46, 80,
139, 240, 416, 720, 1245, 2154, 3728, 6449, 11159, 19307, 33405, 57797, 100000.
}
\label{tab:param}
\end{table}

\section{Observables}
\label{sec:obs}
We will use two replicas in order to create gauge-invariant observables.\cite{Mezard:87}
To identify different replicas we use the superscripts $^{(a)}$ and $^{(b)}$.
The site overlap is defined as
\begin{eqnarray}
 \tau_{\alpha\beta}(\vec{x}) &=& \sigma_{\vec{x}, \alpha}^{(a)}\sigma_{\vec{x}, \beta}^{(b)}\,.\\
 \tau_{\alpha\beta}(\vec{x})^\dag &=& \tau_{\beta\alpha}(\vec{x}) = \sigma_{\vec{x}, \beta}^{(a)}\sigma_{\vec{x}, \alpha}^{(b)}\,.\nonumber
\end{eqnarray}
The order parameter is the overlap tensor:\cite{Fernandez:09}
\begin{equation}
 Q_{\alpha\beta} = \frac{1}{V}\sum_{\vec{x}} \tau_{\alpha\beta}(\vec x) \,.
\end{equation}
This quantity is not rotationally invariant, so we will use the square overlap \cite{Binder:86,Coluzzi:95}
\begin{eqnarray}\nonumber
 Q^2 &=& \mathrm{tr} \left[ Q Q^\dagger \right] \\
     \label{eq:Q2}
     &=&\frac{1}{V^2}\sum_{\vec{x},\vec{y}} \mathrm{tr}\left[\tau(\vec x)\tau(\vec y)^\dag\right] \\
     \nonumber
     &=& \frac{1}{V^2}\sum_{\vec{x},\vec{y}}
	(\vec{\sigma}_{\vec x}^{(a)}\cdot\vec{\sigma}_{\vec y}^{(a)})
	(\vec{\sigma}_{\vec x}^{(b)}\cdot\vec{\sigma}_{\vec y}^{(b)})\,.
\end{eqnarray}
All the $m$ eigenvalues of $Q Q^\dagger $ are rotationally invariant.

Throughout the article, when we will be talking of overlap, we will be referring to the $Q^2$ defined
in Eq.~\eqref{eq:Q2}. The selfoverlap is defined analogously, by taking ${(a)}={(b)}$ in the previous definitions.
Notice that the selfoverlap is not identically equal to 1. It is easy to see, for example, that at
infinite temperature, in the thermodynamic limit it is equal to $Q^2_\mathrm{self}(T=\infty;L=\infty)=1/m$
(see the Appendix).

We will be measuring both point and plane correlation functions. The point correlation function is
\begin{equation}
 C^\mathrm{(point)}(r) = \frac{1}{3V}\sum_{\mu=1}^{3}\sum_{\vec x}^V \mathrm{tr}[\tau(\vec x)\tau(\vec x + \hat e_\mu r)^\dagger]\,,
\end{equation}
where $\mu=1$ (or $x$), 2 (or $y$), 3 (or $z$) is a coordinate axis, and $e_\mu$ is the unitary  vector in that direction.
We also use plane correlation functions because they decay slower and have a better signal-to-noise ratio.
If we denominate the plane-overlap tensor as the mean overlap tensor over a plane
\begin{equation}
 P^x_{\alpha\beta}(x) = \frac{1}{L^2} \sum_{y,z=0}^{L-1} \tau_{{\alpha\beta}}(x,y,z)\,,
\end{equation}
we can define the plane correlation function as
\begin{equation}
 C^\mathrm{(plane)}(r) = \frac{1}{3L}\sum_{\mu=1}^3\sum_{x=0}^{L-1} \mathrm{tr}[P^\mu(x)P^\mu(x+r)^\dagger]\,.
\end{equation}
Both definitions of $C(r)$ are $O(m)$ invariant.
From the correlation functions we measure the second-moment correlation lengths
\begin{align}
\label{eq:xi-punto}
 \xi_2^\mathrm{(point)} &=\sqrt{\frac{\int_0^{L/2} C^\mathrm{(point)}(r) r^4 dr}{\int_0^{L/2} C^\mathrm{(point)}(r) r^2 dr}}\,,\\[2ex]
\label{eq:xi-plano}
 \xi_2^\mathrm{(plane)} &=\sqrt{\frac{\int_0^{L/2} C^\mathrm{(plane)}(r) r^2 dr}{\int_0^{L/2} C^\mathrm{(plane)}(r) dr}}\,.
\end{align}
The difference in the definitions is due to the presence of a Jacobian term when we want to integrate the point correlation function
over the space. These two lengths would be proportional by a factor $\sqrt{6}$ if they had the same purely exponential correlation function.
Note that $\xi_2^\mathrm{(point)}$ and $\xi_2^\mathrm{(plane)}$ are proper estimators of a correlation length only when
the correlation functions $C^\mathrm{(point)}(r)$ and $C^\mathrm{(plane)}(r)$ are connected (i.e. they go to zero for large $r$). 
Otherwise, in principle they could be used to individuate if a quench penetrated in the spin glass phase. In fact, depending on $m$ a quench will drive us
in a ferromagnetic or in a spin glass phase. Our correlation functions are connected in the spin glass phase, but they are 
not in a ferromagnetic state. Consequently, a cumulant such as $\xi_L/L$ will diverge as $L^{\theta/2}$ 
(see Ref.~\onlinecite{Janus:10} for a definition of $\theta$ and an explanation of this behavior) 
when $m$ is too large for a spin glass phase, it will converge as $1/L$ if the quench penetrates in the spin glass phase,
and it will be of order 1 right at the critical $m$, $m_\mathrm{SG}$, that is probably not integer, so not exactly locatable.

When the correlation function decays very quickly and the noise becomes larger than the signal, one could measure
negative values of $C(r)$, that would be amplified by the factors $r^2$ and $r^4$ in the integrals. This would imply
very large errors in $\xi$, or even the square root of a negative number. To overcome this problem, we truncated the
correlation functions when they became less than three times the error.\cite{Belletti:09} This procedure introduces
a small bias, but reduces drastically the statistical error. Furthermore, the plane correlation function required 
the truncation much more rarely, therefore we compared the behaviors as a consistency check.

We also measured the link overlaps
\begin{align}
  Q^2_\mathrm{link}                &= \frac{1}{3V}\sum_{\vec x}^V \sum_{\mu=1}^3 q^{\mu,2}_\mathrm{link} (\vec x) \,, 
  \\  
  \nonumber
  q^{\mu,2}_\mathrm{link} (\vec x) &= \mathrm{tr}\left[ \tau(\vec x)\tau(\vec x + \hat e_\mu)^\dagger\right] = 
  \\[1.5ex]
			           &= 
		    (\vec{\sigma}_{\vec x}^{(a)}\cdot\vec{\sigma}_{\vec x+\hat e_\mu}^{(a)})
		    (\vec{\sigma}_{\vec x}^{(b)}\cdot\vec{\sigma}_{\vec x+\hat e_\mu}^{(b)})\,,
\end{align}
that were shown to be equivalent to the overlaps in the description of the low temperature phase.\cite{Contucci:05,Contucci:06}
The link correlation functions are
\begin{align}
\label{eq:Cpoint}
 C^\mathrm{(point)}_\mathrm{link}(r) &=\frac{1}{9V}\sum_{\mu,\nu=1}^{3}\sum_{\vec x}^V q^{\nu,2}_\mathrm{link} (\vec x)q^{\nu,2}_\mathrm{link} (\vec x+ r\hat e_\mu)\,,\\
\label{eq:Cplane}
 C^\mathrm{(plane)}_\mathrm{link}(r) &= \frac{1}{3L}\sum_{\mu=1}^3\sum_{x=0}^{L-1} P_\mathrm{link}(x)P_\mathrm{link}(x+r)\,,\\[2ex]
 \nonumber \mathrm{with} &  \\
 P^x_\mathrm{link}(x) &= \frac{1}{3L^2} \sum_{\nu=1}^3\sum_{y,z=0}^{L-1} q^{\nu,2}_\mathrm{link}(x,y,z)\,.
\end{align}
The definitions of the link correlation lengths $\xi_\mathrm{2,link}^\mathrm{(point)}$ and $\xi_\mathrm{2,link}^\mathrm{(plane)}$ 
are analogous to Eqs.~\eqref{eq:xi-punto} and \eqref{eq:xi-plano}.

Finally, since we also report measurements of the energy $e(t)$, let us stress that we used the Hamiltonian
${\cal H}$ defined in Eq.~\ref{eq:H}, with a $1/3V$ normalization factor, in order to bound it to one.

Throughout the article, each time one of the defined quantities is referred to the inherent structures
(i.e. the final configurations of our quenches), we
will stress it by putting the subscript $_\mathrm{IS}$.

\section{Results} 
\label{sec:results}
\subsection{Dependence on the number of components}
\label{sec:IS-m}
\begin{figure}[t]
 \includegraphics[width=\columnwidth]{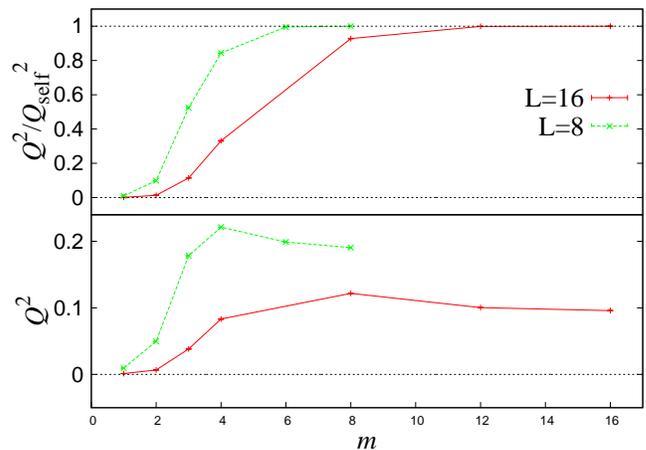}
 \caption{ Dependency of the inherent structures' overlaps from the number of components $m$
 of the spins. The top figure displays the overlap normalized with the self-overlap,
 showing that when $m$ is large enough the energy landscape is trivial. In the bottom
 we have the unnormalized overlap $Q^2$.
 The dashed horizontal lines represent the limits 0 and 1, that bound both observables.
 Error bars are present though small, so not visible.}
 \label{fig:q-m}
\end{figure}

We want to analyze how the model's behavior changes with $m$.
Intuitively, the more components a spin has, the easier it is to avoid frustration \footnote{
Frustration is the impossibility of satisfying all local constraints at the same time.
A way to define it is through the Wilson loop \cite{Parisi:95}. For each closed circuit in the lattice,
we can take the ordered product of all the links that form it. If this product is negative
it is not possible to find a configuration that minimizes simultaneously 
the local energy along each of the links,
and the loop is said to be frustrated \cite{Toulouse:77}. 
When we talk about the system being more or less frustrated we refer to the presence of
a larger or smaller number of frustrated loops.
}\cite{Hastings:00}, and the
simpler is the energy landscape.
According to this scenario, when $m$ increases, the number of available inherent structures decreases
down to the limit in which the energy landscape is trivial, and there is 
only one minimum. 
This should be reflected in the quantity  $Q^2/Q^2_\mathrm{self}$, that should be small when there are many 
minima of the energy, and go to 1
when there is only one inherent structure,
since all the quenches end in the same configuration. 
As shown in 
Fig.~\ref{fig:q-m} (top), our expectation is confirmed. With Ising spins ($m=1$)
the energy landscape is so rich that inherent structures have practically nothing in common.
When we increase $m$ the overlaps start to grow until the limit $Q^2=Q^2_\mathrm{self}$.
By comparing the data for different $L$, we can dismiss a difference in
the behavior between discrete ($m=1$) and continuous ($m>1$) spins, since
$m=1$ for $L=8$ behaves the same as $m=2$ for $L=16$. In Sec.~\ref{sec:dynamics} 
we will discuss aspects in which we do encounter differences. 

Since the number of available inherent structures depends on both $m$
and $L$, we can give an operative definition of a
ratio $(m/L)_\mathrm{SG}$ under which the number of inherent structures is
exponential (so $Q^2/Q^2_\mathrm{self}\simeq0$), and of a ratio $(m/L)_\mathrm{1}$
over which there is only one minimum. 
This way, we can characterize finite-size effects effectively: An extremely 
small system $m/L>(m/L)_\mathrm{1}$ is trivial and has only one stable state. Increasing
the size we encounter a less trivial behavior, but to find a visible signature 
of a spin glass phase one has to have $L\geq m (L/m)_\mathrm{SG}$. From Fig.~\ref{fig:q-m}
one can see that for $L=8$, $m_\mathrm{SG}=1$, and for $L=16$, $m_\mathrm{SG}=2$. 
Then, for example, we see that to observe a complex behavior for $m=3$
spin glasses, one should use $L>16$.

Moreover, this interpretation gives a straightforward explanation of the finite-size effects
one encounters in the energy of an inherent structure (Table~\ref{tab:IS-m}). For example,
if we compare $L=8,16$ at $m=8$, we notice two incompatible energies. 
In fact, there is an intrinsic difference between the two
sizes, since $L=8$ represents single-basin systems, while $L=16$ has a variety of inherent
structures.
On the other side, finite-size effects on lower $m$ are smaller, because we are comparing
similar types of behavior.

\begin{figure}
 \includegraphics[width=\columnwidth]{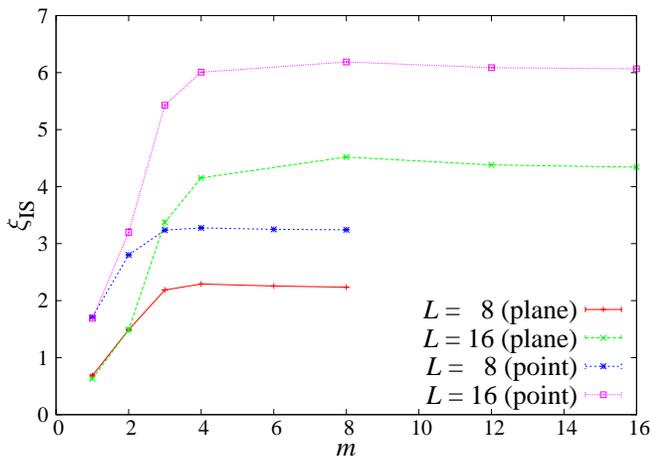}
\caption{Dependency of the second-moment correlation length $\xi_2$
on the number of components of the spins $m$. We show both the plane and the point
correlation functions defined in Eqs.~\eqref{eq:Cpoint} and \eqref{eq:Cplane}, for $L=8,16$.}
\label{fig:xi-m}
\end{figure}

Notice that, although the ratio  $Q^2/Q^2_\mathrm{self}(m)$ grows 
monotonously, this is not true for the pure overlap $Q^2(m)$ (Fig.~\ref{fig:q-m}, 
bottom), that has a peak at an intermediate $m$. Moreover, the position of the peak
doubles when we double the lattice linear size, justifying the operational definitions
$(m/L)_\mathrm{SG}$ and $(m/L)_\mathrm{1}$. 
The same peak at intermediate $m$ is also visible in the energy and in the correlation
length (Fig.~\ref{fig:xi-m}), indicating that there is an intrinsic difference
in the nature of the reached inherent structures. 
In Table~\ref{tab:IS-m} we give the values
of the aforementioned observables at the inherent structure.
\begin{table}[b]
\footnotesize
\begin{ruledtabular} 
 \begin{tabular}{ccccccc}
  \normalsize{$L$} & \normalsize{$m$} & \normalsize{$e_\mathrm{IS}$} & \normalsize{$Q^2_\mathrm{IS}$} & \normalsize{$Q^2_\mathrm{self,IS}$}  & \normalsize{$\xi^\mathrm{(plane)}_\mathrm{IS}$}& \normalsize{$\xi^\mathrm{(point)}_\mathrm{IS}$} \\\hline\hline
  8   & 1  & -0.4709(1)   & 0.0095(1)  &  1           &  0.68(2)   &  1.71(1) \\
  8   & 2  & -0.5953(1)   & 0.0497(3)  &  0.50297(2)  &  1.49(1)   &  2.802(4) \\
  8   & 3  & -0.6151(1)   & 0.1784(6)  &  0.33994(4)  &  2.188(2)  &  3.2358(7) \\
  8   & 4  & -0.6176(2)   & 0.2213(5)  &  0.26229(9)  &  2.2919(9) &  3.2760(5) \\
  8   & 6  & -0.61801(11) & 0.1989(1)  &  0.1997(1)   &  2.2567(3) &  3.2514(2) \\
  8   & 8  & -0.61797(12) & 0.1905(1)  &  0.1905(1)   &  2.2364(3) &  3.2428(2) \\\hline
 16   & 1  & -0.4721(1)   & 0.00123(6) &  1           &  0.63(2)   &  1.69(1) \\
 16   & 2  & -0.5965(1)   & 0.0067(2)  &  0.500379(8) &  1.49(4)   &  3.20(6) \\
 16   & 3  & -0.6165(1)   & 0.0382(5)  &  0.33416(1)  &  3.37(3)   &  5.43(1) \\
 16   & 4  & -0.6191(2)   & 0.0833(6)  &  0.25144(2)  &  4.153(7)  &  6.008(4)	 \\
 16   & 8  & -0.6200(1)   & 0.1218(3)  &  0.13126(5)  &  4.519(2)  &  6.187(1) \\
 16   & 12 & -0.6202(1)   & 0.10031(9) &  0.10044(9)  &  4.3814(8) &  6.087(1) \\
 16   & 16 & -0.6197(1)   & 0.0959(1)  &  0.0959(1)   &  4.3412(8) &  6.066(1)\\\hline
 64   & 3  & -0.61657(4)  & 0.00064(2) &  0.3333466(4)&  3.53(7)   &  6.74(6) \\
\end{tabular}
\end{ruledtabular}
\caption{Properties of the inherent structures. For each choice of the parameters
we show the observables at the end of the quench: The energy $e_\mathrm{IS}$, the 
overlap $Q^2_\mathrm{IS}$, the selfoverlap $Q^2_\mathrm{self,IS}$, the point-correlation 
length $\xi^\mathrm{point}_\mathrm{IS}$ and the plane correlation length 
$\xi^\mathrm{plane}_\mathrm{IS}$.
}
\label{tab:IS-m}
\end{table}
We see in this behavior the competition between two effects.
When $m$ is small, the quench has a vast choice of valleys where to fall. 
Since, reasonably the attracion basin of the lower-energy inherent structures is larger, 
the wide variety of inherent structures will increase the probability of
falling in a minimum with low energy and larger correlations.
When $m$ increases, the number of available valleys decreases, so it is
more likely that two different replicas fall in the same one. Yet, the \emph{quality}
of the reached inherent structures decreases, since the quench does not have the
possibility to choose the lowest-energy minimum.

\subsection{Overlap Probability Densities}
\begin{figure}[t]
 \includegraphics[width=\columnwidth]{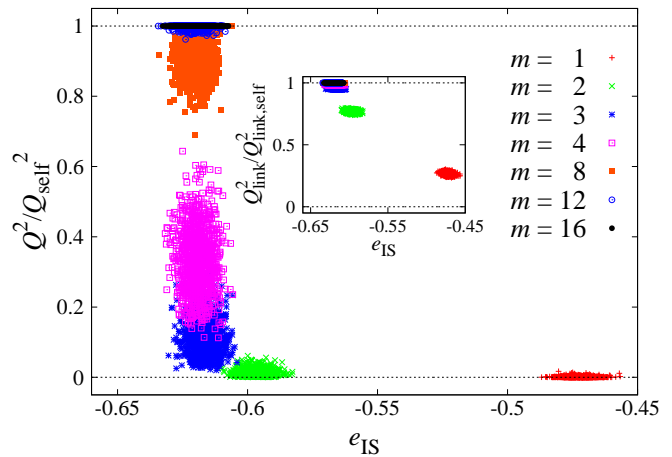}
 \caption{Scatter plots for $L=16$, at different values of $m$, of
 the overlap ratio  $Q^2_\mathrm{IS}/Q^2_\mathrm{self,IS}$ against
 mean energy between the two replicas $e_\mathrm{IS} = (e_\mathrm{IS}^{(a)}+e_\mathrm{IS}^{(b)})/2$.
 Each simulated sample contributes to the plot with a single point.
  The inset displays an analog plot for the link overlap.}
\label{fig:Eq}
\end{figure}

From these observations it is reasonable to think that overlap and energy of the inherent structures
are correlated. We looked for these correlations both on the overlap, on the selfoverlap,
and in their ratio, but with a negative result. In Fig.~\ref{fig:Eq} we show a scatter-plot
of the ratio of the inherent structure's overlaps $Q^2_\mathrm{IS}/Q^2_\mathrm{self,IS}$
that confirms our statements.

\begin{figure}[b]
 \includegraphics[width=\columnwidth]{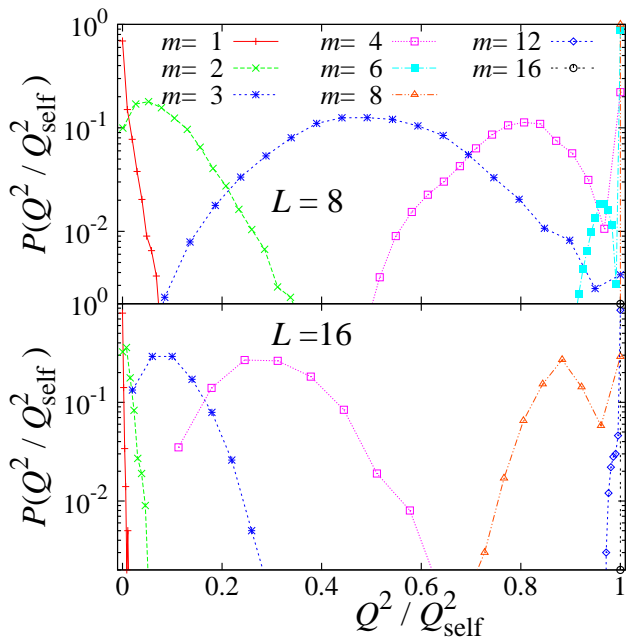}
 \caption{Overlap probability distribution functions of 
  the inherent structures for different values of $m$. The top figure depicts
  data for $L=8$, on the bottom we have $L=16$. 
  The curves are normalized to plot all the curves together. The actual probability distribution
  function is obtained by dividing each point by the bin width $\Delta Q/N_\mathrm{bins}$, where 
  $\Delta Q$ is the difference between maximum and minimum $Q^2$.}
 \label{fig:Pq}
\end{figure}

\begin{figure}[t]
 \includegraphics[width=0.485\columnwidth]{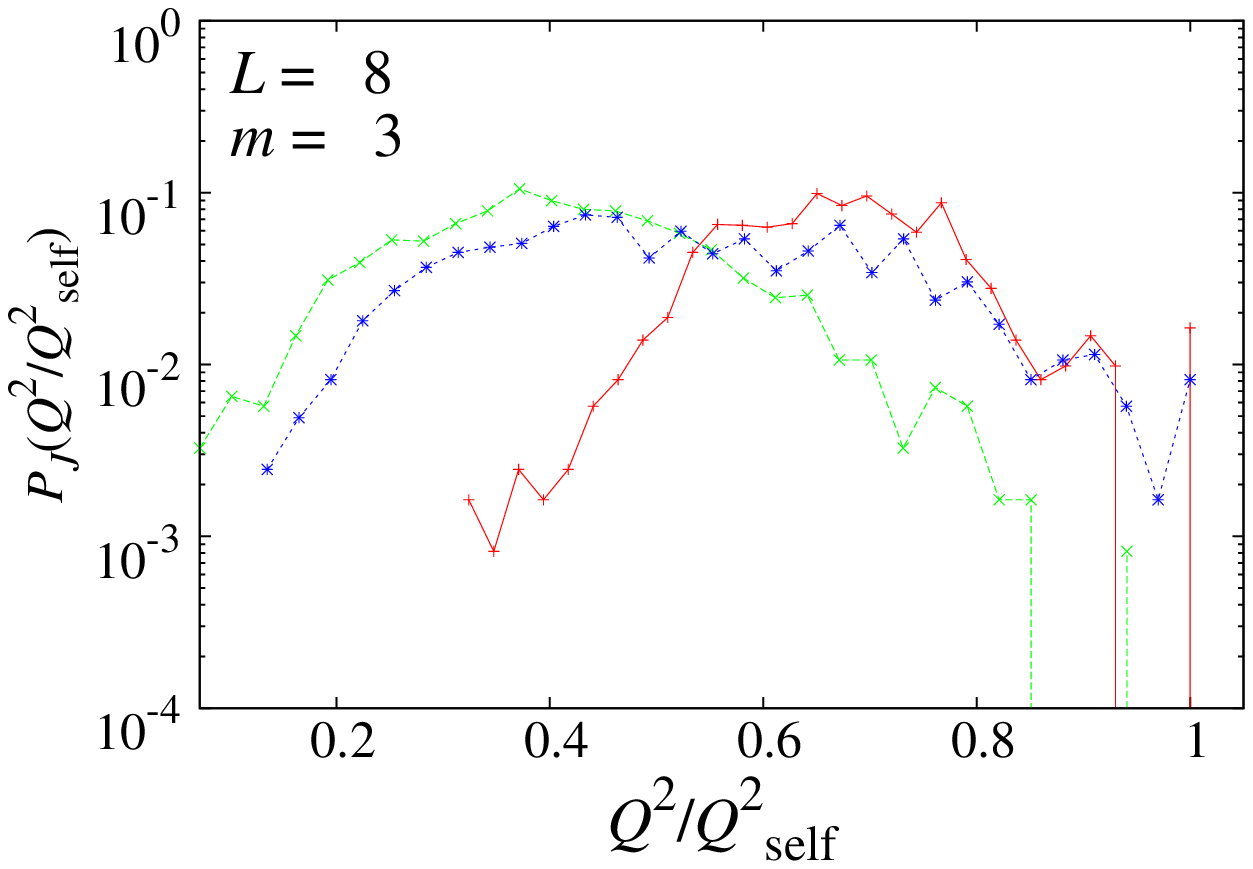}
 \includegraphics[width=0.485\columnwidth]{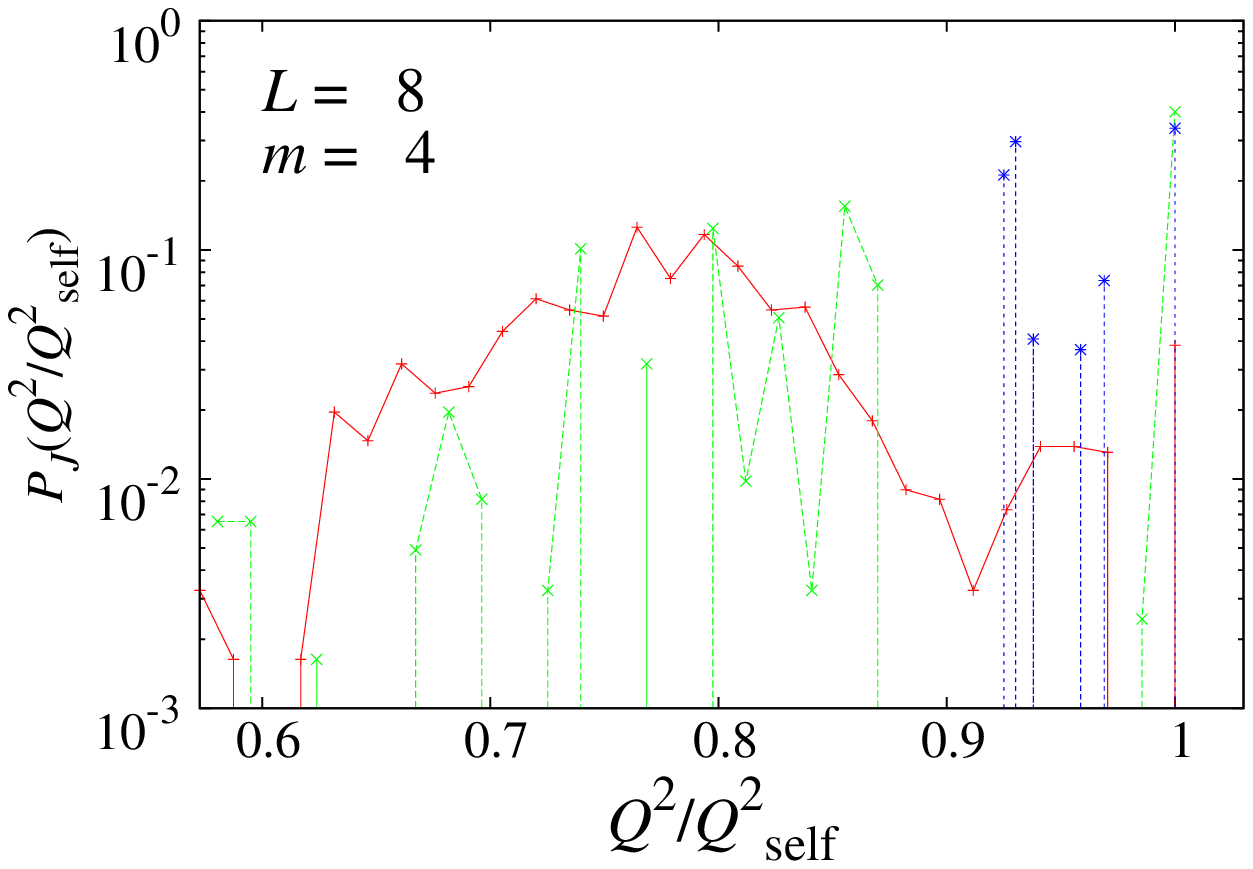}
 \includegraphics[width=0.485\columnwidth]{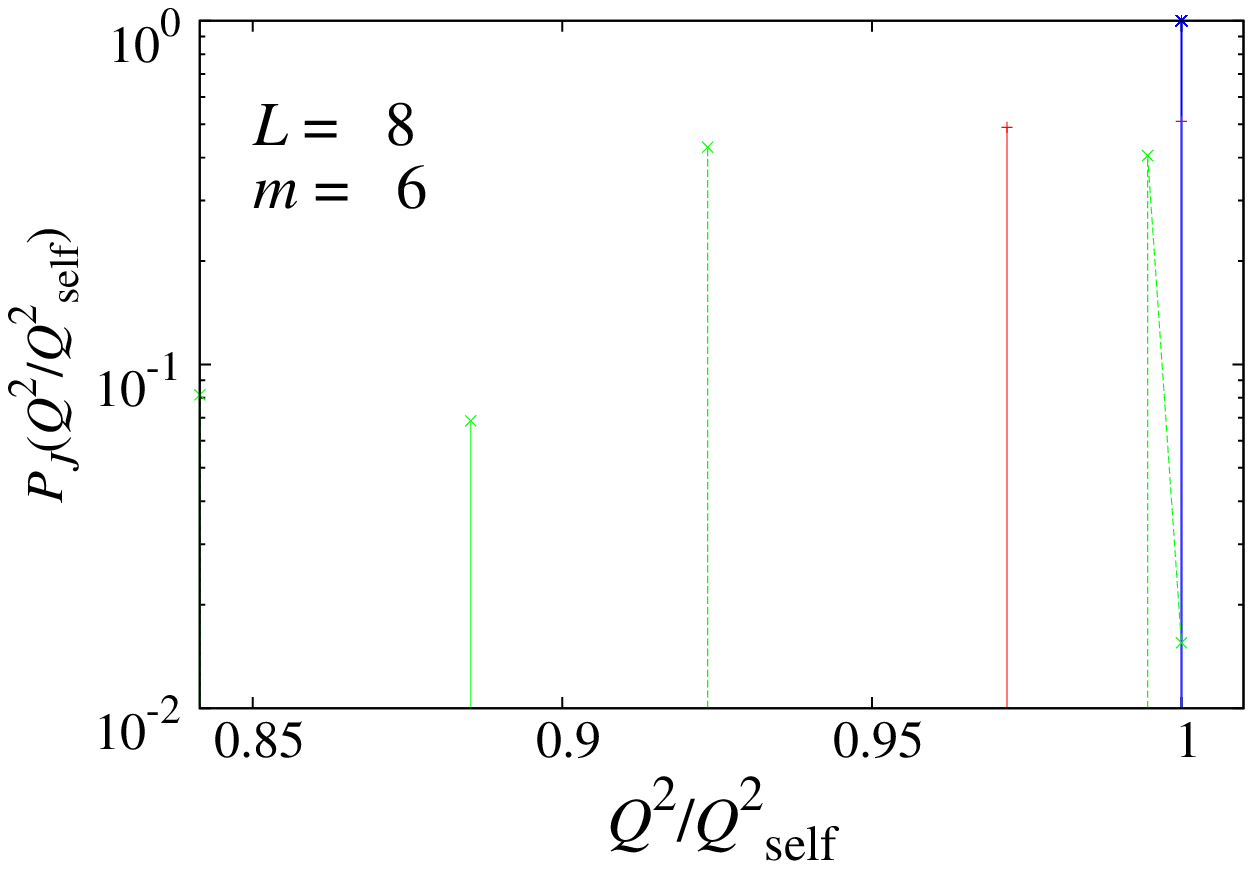}
 \includegraphics[width=0.485\columnwidth]{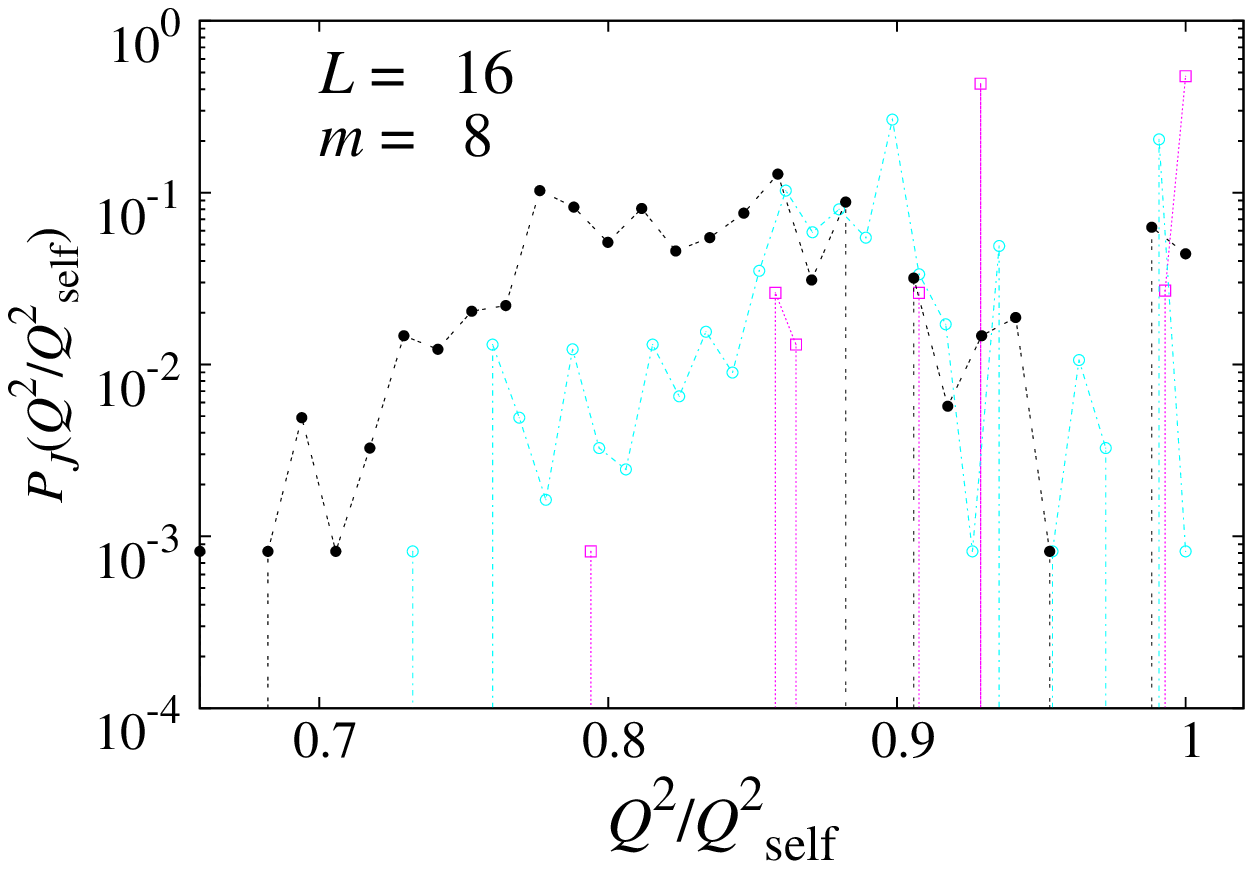}
 \caption{Sample-dependent overlap probability distribution functions $P_{J}(Q^2_\mathrm{IS}/Q^2_\mathrm{self,IS})$. 
 Each curve depicts data from a separate sample. 
 In each plot we show a selection of three samples with different shapes of the distribution. 
 The choices of the parameters are represented in the key of each plot. 
 We used two different color codes to distinguish the three plots that come from $L=8$ systems 
 (top-left and right, and bottom left), from the bottom-right plot
 that is for $L=16$.
  The curves are normalized as in Fig.~\ref{fig:Pq}.}
 \label{fig:PJq}
\end{figure}

The cross sections of Fig.~\ref{fig:Eq} give an idea of the energy and overlap probability
distribution functions. We show explicitly the overlap probability distribution
functions (normalized with the bin width) of the inherent structures in Fig.~\ref{fig:Pq}.
They are qualitatively different from their thermal counterparts (see, e.g., 
Ref.~\onlinecite{Janus:10}). The ratio $Q^2_\mathrm{IS}/Q^2_\mathrm{self,IS}$ is bounded between 
zero and one. The distributions are extremely wide, and the phenomenology is 
quite different near the two bounds. In fact, when $m$ is large enough,
the limit $Q^2_\mathrm{IS}/Q^2_\mathrm{self,IS}=1$ changes completely the shape of the curves, 
introducing a second peak (that we could read as an echo of the Bose-Einstein condensation
remarked in Ref.~\onlinecite{Aspelmeier:04}).
Around the lower bound of the $P(Q^2_\mathrm{IS}/Q^2_\mathrm{self,IS})$, instead, there is no double
peak.
We can try to give an interpretation to the presence of this second peak by looking at the overlap
distribution functions $P_J(Q^2_\mathrm{IS}/Q^2_\mathrm{self,IS})$ for a given instance of the
couplings. In Fig.~\ref{fig:PJq} we show that this distribution has relevant sample-to-sample fluctuations.
When we increase $m$, the number of minima of the energy, $N_\mathrm{IS}$, gradually becomes smaller. Yet,
depending on the specific choice of the couplings, $N_\mathrm{IS}$ can vary sensibly. For example in Fig.~\ref{fig:PJq}, top-right,
one can see that when $L=8$ and $m=4$, $N_\mathrm{IS}$ can be both large (red curve) or
of order one (blue curve). For $L=8$, $m=6$ (Fig.~\ref{fig:PJq}, bottom-left), the situation is similar: for the blue curve $N_\mathrm{IS}=1$,
while for others $N_\mathrm{IS}>1$.

As we similarly stated in Sec.~\ref{sec:IS-m}, we notice that the lattice size plays
a substantial role on the properties of the reached inherent structure, since when we pass from
$L=8$ to $L=16$ histograms regarding the same $m$ cover very different ranges of $q$. We can both
see them traditionally as strong finite-size effects, or focus on $L$ as a relevant parameter (as it was
suggested, for example, in Ref.~\onlinecite{Baityjesi:14}), concentrating the interest on finite $L$.

\subsubsection{Link Overlaps}
\begin{figure}[t]
 \includegraphics[width=\columnwidth]{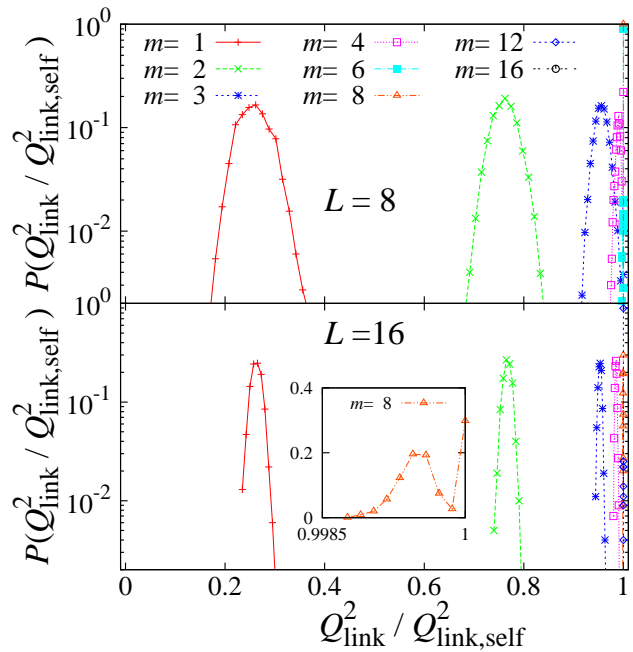}
 \caption{Same as Fig.~\ref{fig:Pq}, but for the link overlap. The inset shows a zoom for the $m=8$, $L=16$ data,
where we also removed the logarithmic scale on the $y$ axis.
 }
 \label{fig:Pqlink}
\end{figure}

Since in the past ten years an increasing attention has been devoted to the link overlap $Q^2_\mathrm{link}$
as an alternative order parameter for the study of the low temperature region of spin
glasses,\cite{Krzakala:00,Contucci:06,Janus:10} in Fig.~\ref{fig:Pqlink} we show also the link-overlap 
histograms $P(Q^2_\mathrm{link,IS}$) at the inherent structures.
The functions $P(Q^2_\mathrm{link,IS})$ have much smaller finite-size effects than the $P(Q^2_\mathrm{IS})$,
and are more Gaussian-like (although the Gaussian limit is impossible, since $Q^2_\mathrm{link}$ is bounded
between 0 and 1). The inset shows that the second peak on high overlaps is present also
with the link overlap.

\begin{figure}[!ht]
\centering
 \includegraphics[width=0.48\columnwidth]{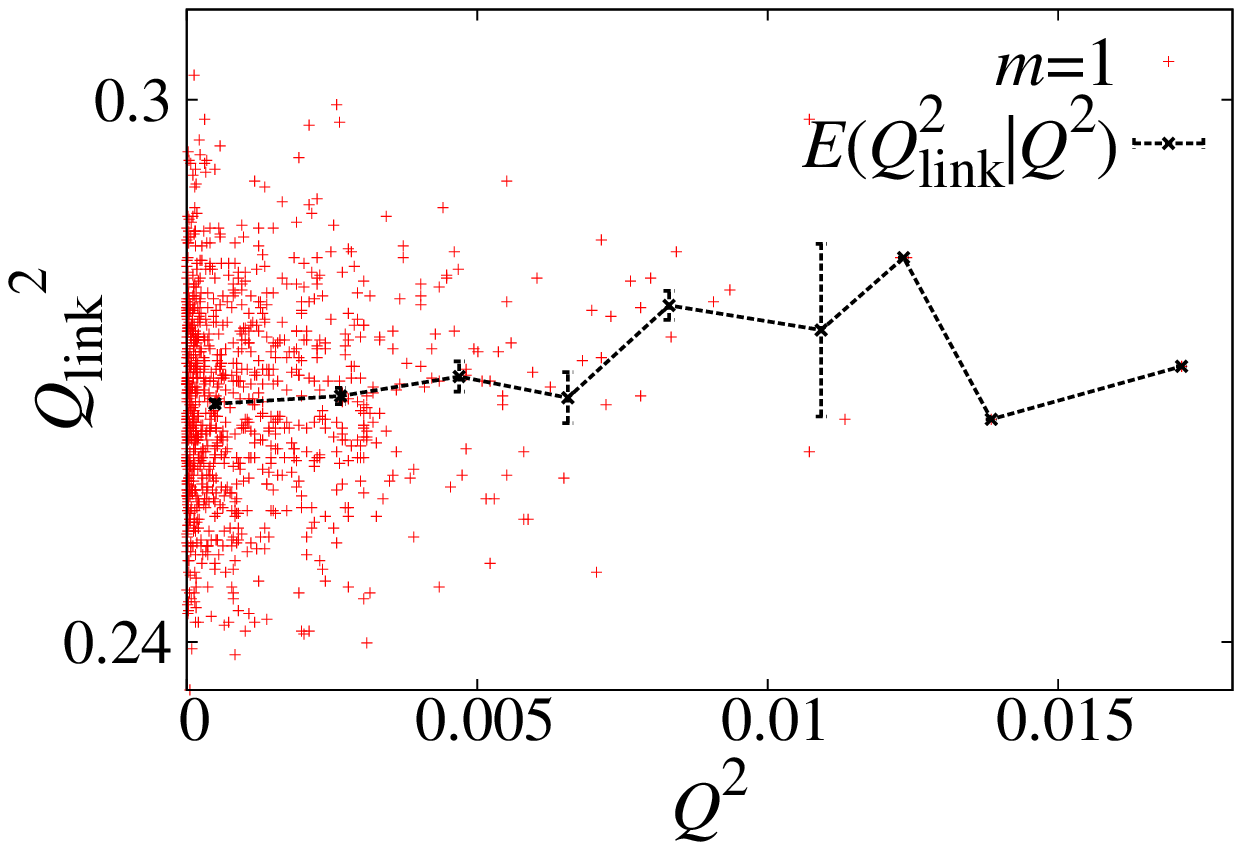}
 \includegraphics[width=0.48\columnwidth]{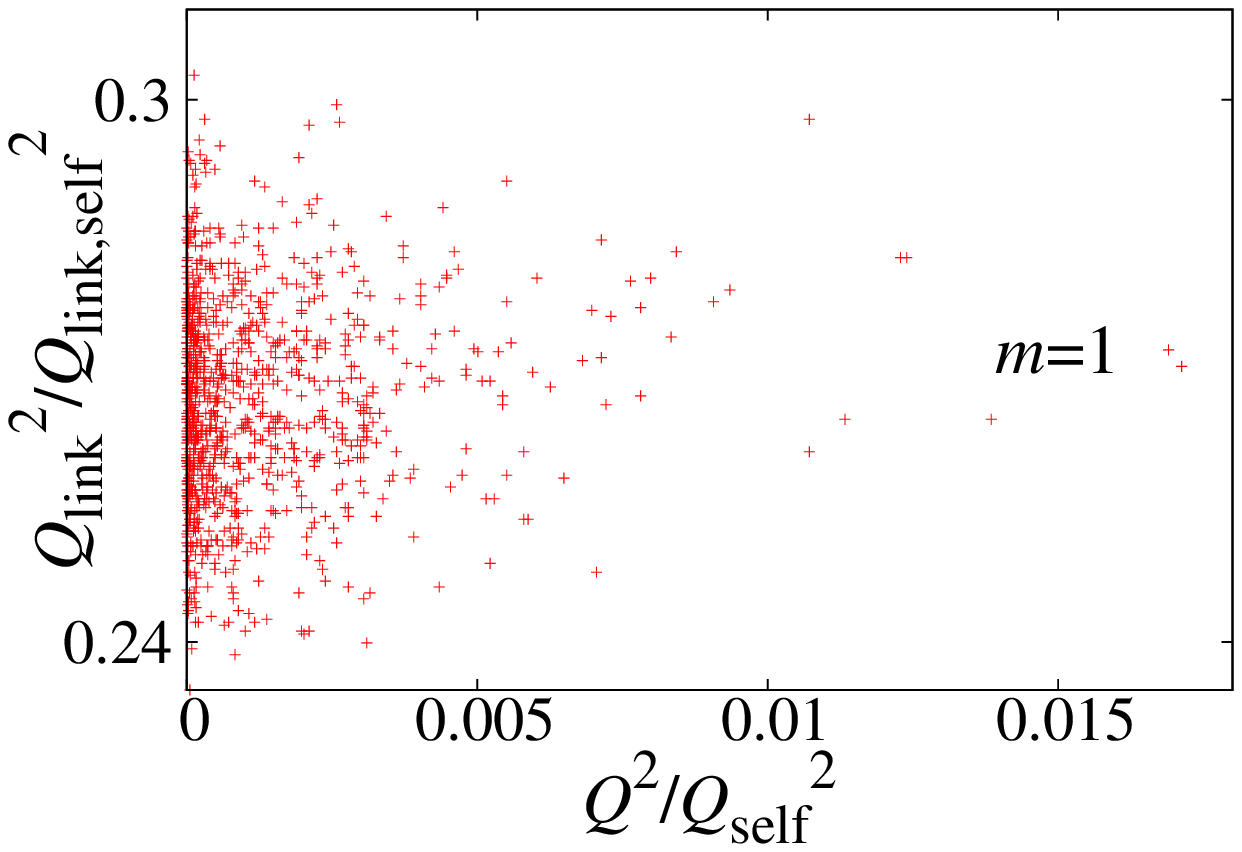}
 \includegraphics[width=0.48\columnwidth]{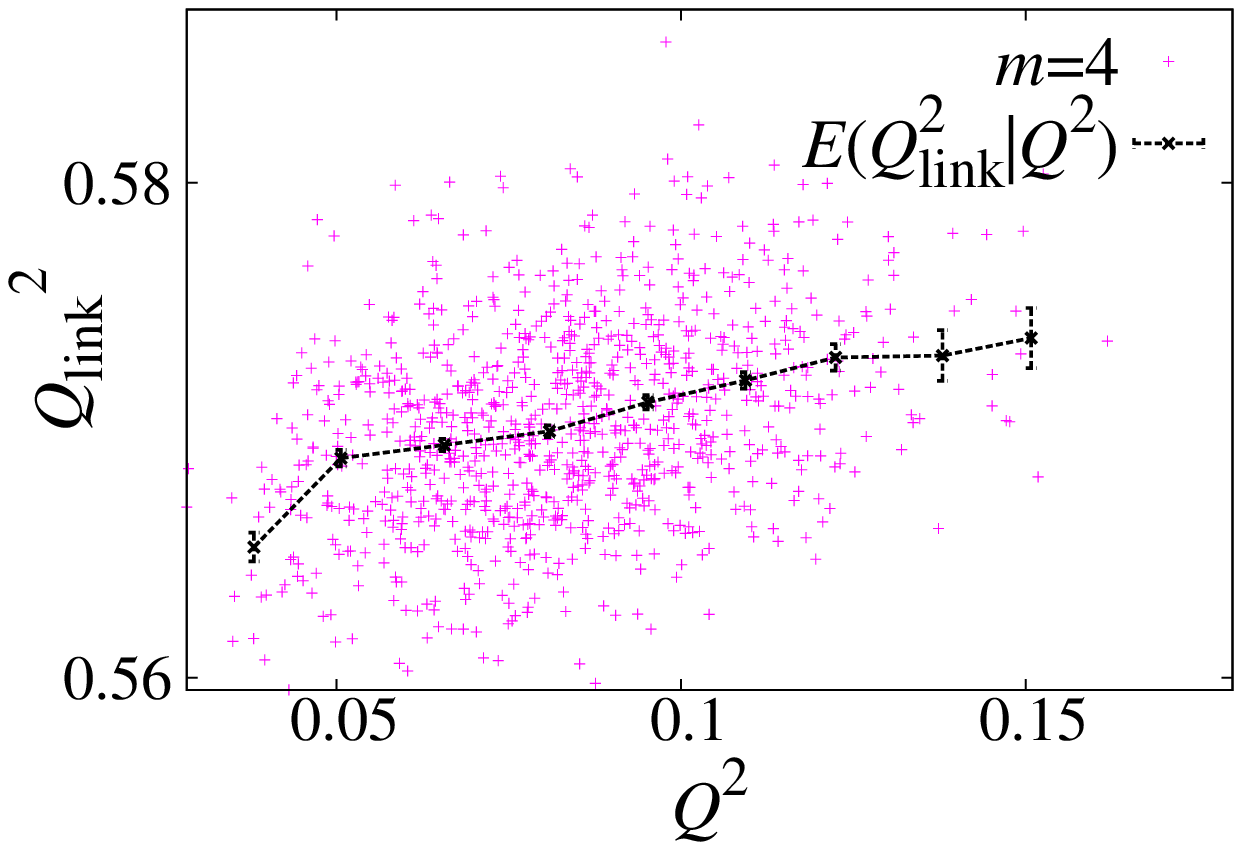}
 \includegraphics[width=0.48\columnwidth]{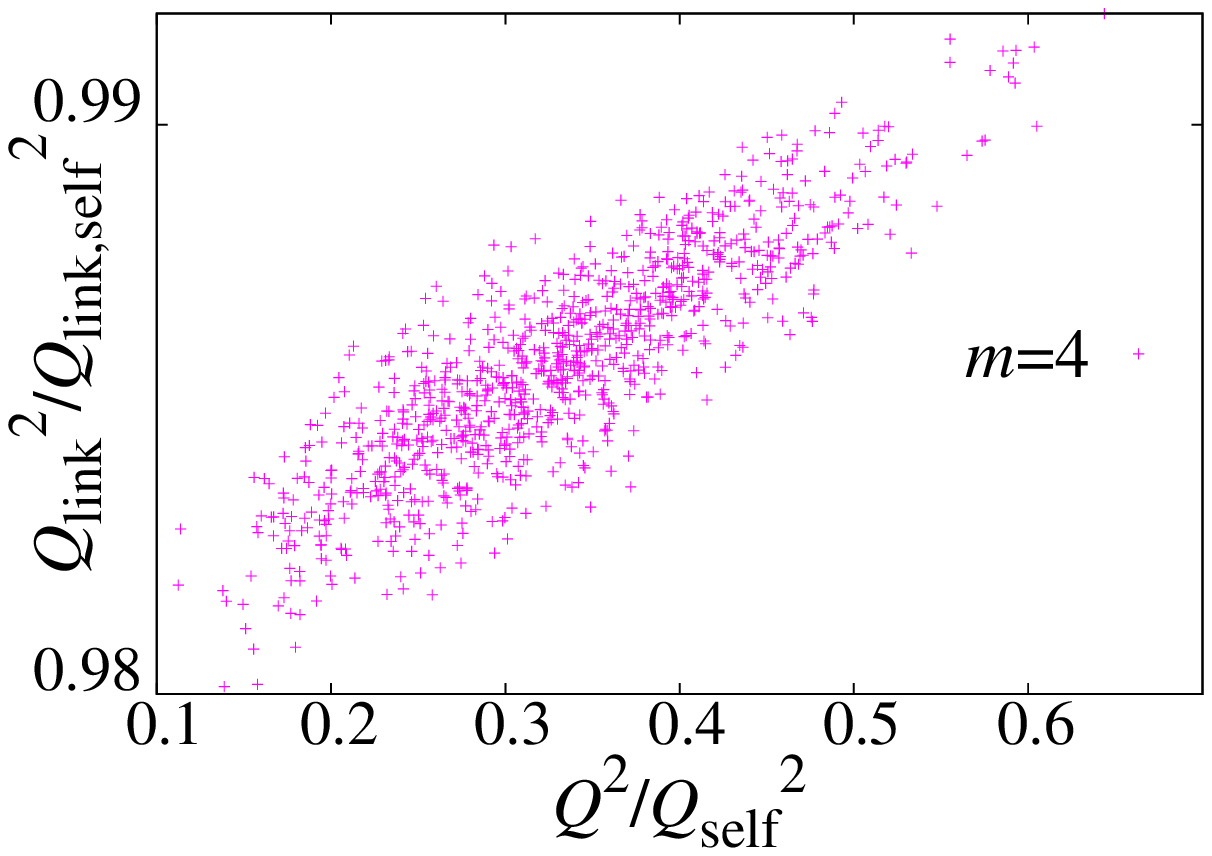}
 \includegraphics[width=0.48\columnwidth]{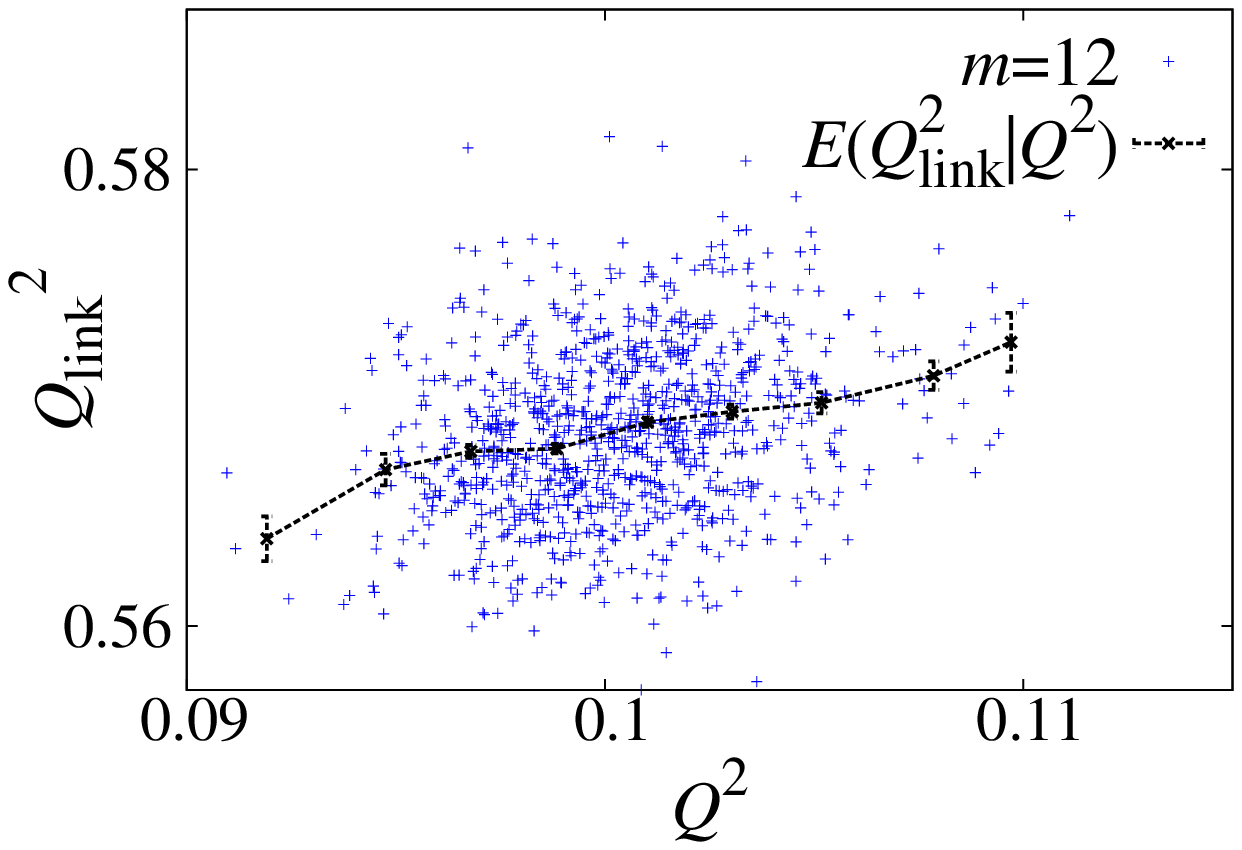}
 \includegraphics[width=0.48\columnwidth]{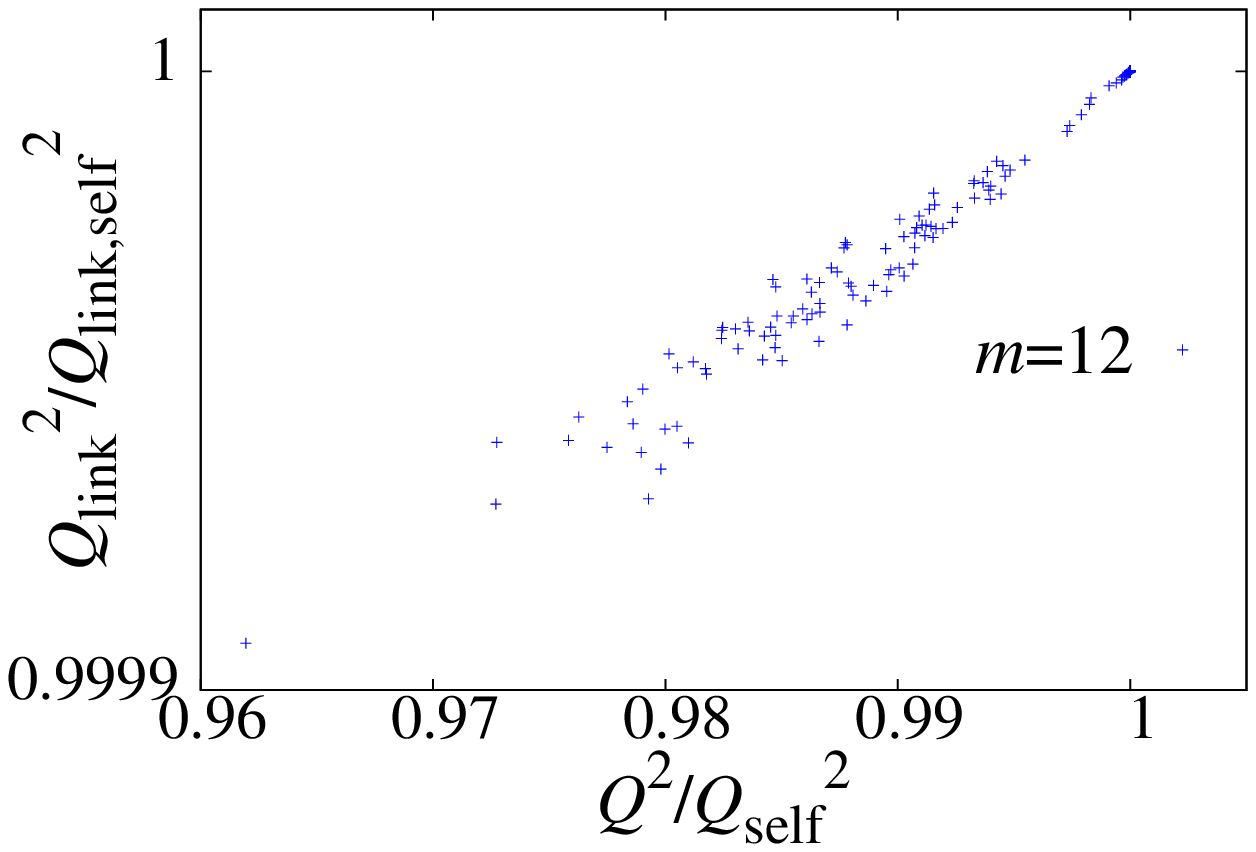}
 \caption{Correlation between the spin and the link overlap of the inherent structures, 
 for $L=16$ lattices, with $m=1$ (top), $m=4$ (center) and 
 $m=12$ (bottom). On the left we plot the overlaps, while on the right they
 are normalized with the self overlap. Normalizing with the self overlap
 increases the correlations between the two order parameters. The two top figures are the same
 because the self overlap is one when $m=1$.
 The black lines on the left plots represent $E(Q_{link}^2|Q^2)$, and they show that a correlation 
 exists also without normalization.}
 \label{fig:scatterplot_qqlink}
\end{figure}

We checked also the correlation between spin and link overlaps. At finite temperature there are
different predictions between RSB and droplet pictures. According to the RSB picture the conditional
expectation value $E(Q^2_\mathrm{link}|Q^2)$ should to be a linear, strictly increasing function of
$Q^2$, while this should not be true in the Droplet theory.
When $m$ is small, this correlation is practically invisible, but it becomes extremely strong when
we increase the number of components of the spins (Fig.~\ref{fig:scatterplot_qqlink}). 
Notice how the correlation between spin and link overlap is formidably increased when we normalize
the two with the selfoverlap.
The curves in Fig.~\ref{fig:scatterplot_qqlink} represent $E(Q^2_\mathrm{link}|Q^2)$.
If we exclude the tails, that are dominated by rare non-Gaussian events, the trend is 
compatible with linearly increasing functions.

\subsection{Quench Dynamics}
\label{sec:dynamics}
\begin{figure}[t]
 \includegraphics[width=\columnwidth]{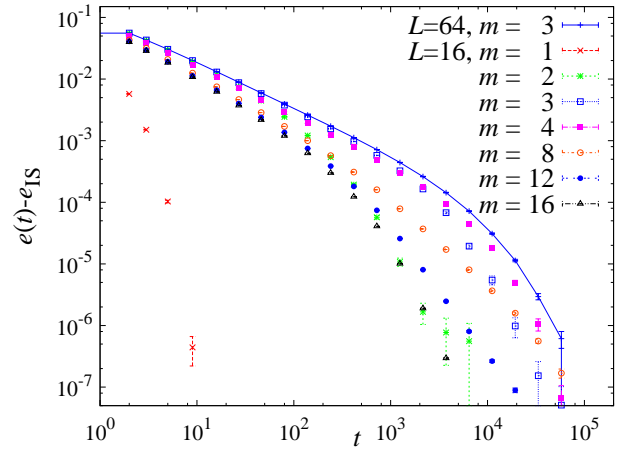}
  \caption{ Evolution of the energy during the quench for all the simulated values of $m$,
  in $L=16$ lattices. On the $x$ axis there is the time, measured in full lattice quench sweeps. On the
  $y$ axis there is the difference between the energy at time $t$, $e(t) = \left(e^{(a)}(t)+e^{(b)}(t)\right)/2$, and its final value $e_\mathrm{IS} = e(t=10^5)$.
  The convergence speed is very different between continuous and discrete spins.
  To stress the finite-size effects we also show points for $L=64$, $m=3$ (points connected by segments).}
 \label{fig:evol-E}
\end{figure}

Let us get an insight on the dynamics of the quench.
For short times, the energy converges towards a minimum with a roughly power law behavior (Fig.~\ref{fig:evol-E}). 
At longer times there is a cutoff, that grows with the system's size,
revealing a change in the dynamics after which the system converges faster to a valley.
We stress the great difference in the 
convergence rate between $m=1$ and $m>1$. We can identify two different decrease rates, depending on
whether the spins are discrete or continuous.

\begin{figure}
\includegraphics[width=\columnwidth]{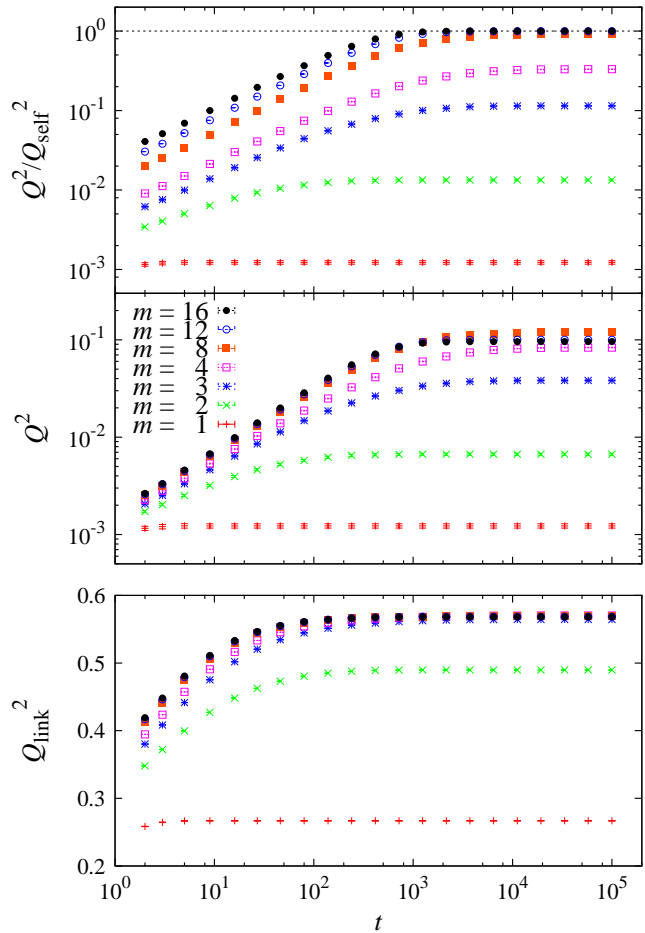}
 \caption{ Time evolution of the overlaps in $L=16$ lattices. 
  In the top set we show the overlap $Q^2$ normalized with the selfoverlap $Q^2_\mathrm{self}$.
  On the center we show $Q^2$ without normalizing. Notice that differently from the top case,
  in the bottom plot it is the curve representing $m=8$ that reaches the highest values.
  The third plot shows that the behavior is analogous with $Q^2_\mathrm{link}$.
  }
 \label{fig:evol-q}
\end{figure}

Fig.~\ref{fig:evol-q} shows the evolution of the overlap for $L=16$, and gives a better understanding of why quantities
such as $Q^2_\mathrm{IS}$ are not monotonous with $m$. We show both the evolution of $Q^2/Q^2_\mathrm{self}$ (top),
and of $Q^2$ (bottom). The first one behaves as one would expects when the number of minima is decreasing to one. 
On the other side, we see from the lower plot
\emph{how} the quenches of $m=8$ reach the highest overlap. 
A possible interpretation is to ideally separate the quench in two regions. At the beginning 
there is a search of the valley with a power-law growth of $Q^2$, and later the convergence inside of the valley. 
Fig.~\ref{fig:evol-q} shows that the search of the valley stops earlier when $m=12,16$, i.e. when their number is
of order one.

\begin{figure}
 \includegraphics[width=\columnwidth]{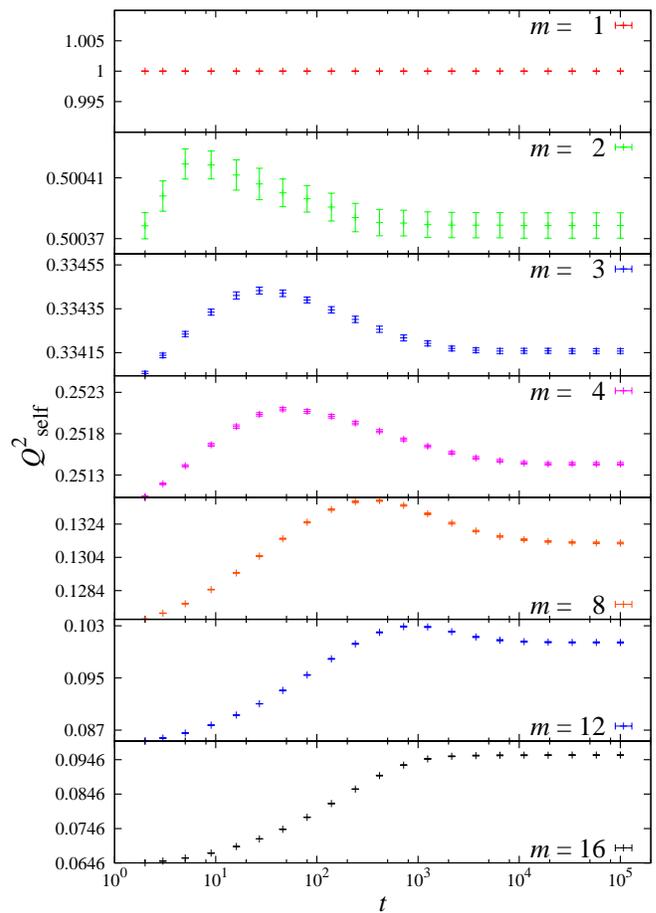}
  \caption{Evolution of the selfoverlap $Q^2_\mathrm{self}(t)$ for lattices of size $L=16$, for different values of $m$.
  Note the differences in the $y$-scales: For small $m$ the variation of $Q^2_\mathrm{self}(t)$ is very small, while
  for the largest ones it is of the order of the self-overlap.}
 \label{fig:evol-qself}
\end{figure}

We remark on a nonlinear trend on the evolution of the selfoverlap $Q^2_\mathrm{self}(t)$. For continuous spins 
($m>1$) it has a different value at infinite and zero temperature (Fig.~\ref{fig:evol-qself}). 
This variation is strikingly visible when $m$ is large,
but the same trends are found for $m\leq3$, though the variations are so small that it is justified that they are usually not found.
\footnote{To our knowledge, the only reference where a non-trivial behavior of the self-overlap was found is in Ref.~\onlinecite{Marco:11}.
Yet, in this case it was in the study of inherent structures from finite temperature, and in the chiral sector (they worked with $m=3$).}
Moreover $Q^2_\mathrm{self}(t)$ is highly nonlinear, and, except for the highest $m$, it overshoots before having converged.

\begin{figure}[t]
 \includegraphics[width=\columnwidth]{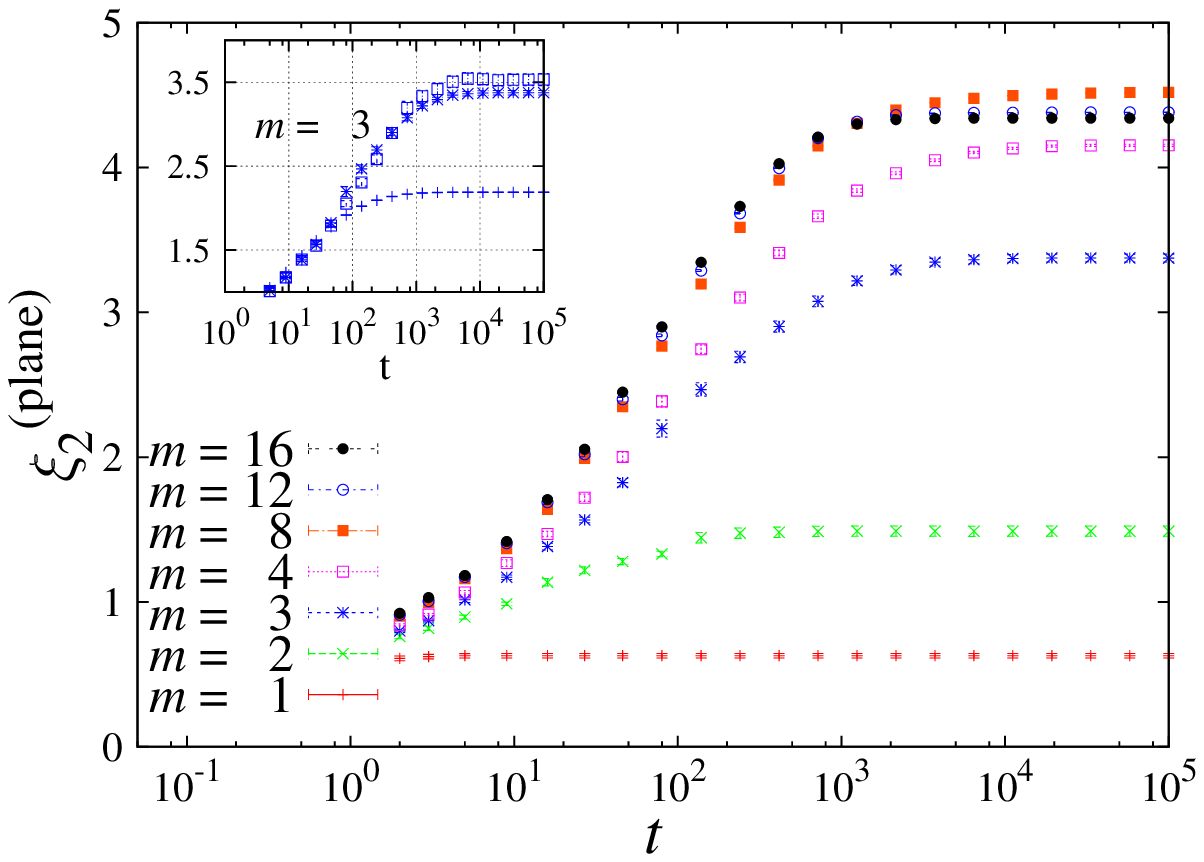}
\caption{Time evolution of the plane second-moment correlation length $\xi_2^\mathrm{plane}$.
In the large figure we show every simulated $m$ for size $L$. Notice that the highest correlation
length is reached by $m=8$. The inset depicts the sole case of three-dimensional spins ($m=3$) for
sizes $L=8,16,64$.}
\label{fig:evol-xi}
\end{figure}

In Fig.~\ref{fig:evol-xi} we show the evolution of the correlation lengths $\xi_2^\mathrm{plane}$ during the quenches for $L=16$
for all our values of $m$. 
We see the same variety of behaviors shown by $Q^2$ (Fig.~\ref{fig:evol-q}), with $\xi_2^\mathrm{plane}(m=12,16)$
that abruptly stop increasing, while when $m=8$ the increase is similar but lasts longer and the change of growth is smoother.
\footnote{The point correlation length $\xi_2^\mathrm{point}$ behaves analogously.}

We can contrast our results with the ones obtained by Berthier and Young in Ref.~\onlinecite{Berthier:04}
for $m=3$ Heisenberg spin glasses. In that case they measured the evolution of the coherence
length in quenches down to positive temperature $T_0>0$ ($L=60$).
They remarked two different regimes of growth of the coherence length, and attributed them to the 
passage from critical to activated dynamics \footnote{Note that the definition of the coherence length in Ref.~\onlinecite{Berthier:04} is
different from ours.}.
In that case the slope of the second phase kept being positive
and $\xi$ did not appear to converge after $10^5$ lattice sweeps.
We can make a direct comparison with our quenches to zero-temperature $T_0=0$ with $L=64$ (Fig.~\ref{fig:evol-q}, inset).
We obtain a flat second regime after $10^4$ sweeps, so we can indeed attribute the growth in the 
second regime to thermal effects.
In the inset we compare the coherence length of different lattice sizes to remark 
that although 
$\xi_2^\mathrm{plane}<4$, we are clearly far from the thermodynamic limit even for $L=16$.

\section{Conclusions}
\label{sec:conc}
We performed an extensive study of the energy landscape of three-dimensional vector spin glasses, focusing
on their dependence on the number of components $m$ of the spins. We were concerned both
with the zero-$T$ dynamics and with the properties of the inherent structures, remarking
various types of finite-size effects.

Increasing $m$ the number of minima in the energy landscape decreases monotonously,
down to the limit of a single state. The number of components $m_\mathrm{SG}(L)$ after which the 
number of minima becomes subexponential grows with the lattice size. Reversing the relation, 
we can operatively define $L_\mathrm{SG}(m_\mathrm{SG})$ as the smallest lattice size needed
in order to observe a complex behavior for a given $m$.

For small $m$ correlations are small and dynamics are trivial, while when $m$ becomes
larger correlations increase and the convergence to an inherent structure slows down (for a small enough $m/L$ ratio).
We remark on the competition between the $m=1$ limit, with abundance of inherent structures,
and the large-$m$ limit where at $T=0$ there is only a single state. 

In finite systems neither the overlap, nor the correlation length, nor the energy
of the inherent structures is a monotonous function of $m$, as one would expect from a decreasing number
of available disordered states. They have instead a peak at an intermediate $m$.
We attribute this to the fact that when there are several minima, those of more ordered states have a larger
attraction basin, so having many inherent structures makes it easier to fall into a more ordered state.
If one wanted to rule out the non-monotonous behavior it could be useful to redefine the correlations
as a function of the normalized overlaps $Q^2/Q^2_\mathrm{self}$, as we have seen that the normalized overlaps
do exhibit a monotonous trend.

Also, we presented probability distribution functions (pdf)
of the spin and link order parameters $Q^2/Q^2_\mathrm{self}$ and $Q^2_\mathrm{link}/Q^2_\mathrm{link,self}$, 
noticing that the states with $Q^2/Q^2_\mathrm{self}=1$ have a major attraction basin,
and create a second peak in the curve. Finite-size effects in the inherent structures' pdfs
were very heavy, as remarked also by looking at other observables, but they were minimal
if we considered the link overlap. Also, the dependency between $Q$ and $Q_\mathrm{link}$
is consistent with RSB predictions.

Finally, we found a non-trivial behavior on the evolution of the self-overlap, 
that could be used as an indicator of the ``quality'' of a reached inherent structure.


\section*{Acknowledgements}
We thank V\'ictor Mart\'in-Mayor for useful conversations and a careful reading of the manuscript.

We were supported by the European Research Council under the European
Union’s Seventh Framework Programme (FP7/2007-2013, ERC grant agreement 
no. 247328). 
We were partly supported by MINECO, Spain,
through the research contract No. FIS2012-35719-C02.
This work was partially supported by the GDRE 224 CNRS-INdAM GREFI-MEFI.
M.B.-J. was supported by the FPU program (Ministerio de Educaci\'on, Spain).

\appendix

\section{Self overlap at infinite temperature}
\label{app:selfoverlap}
We show here the simple case of dependence on $m$ of the selfoverlap $Q^2_\mathrm{self}$ 
at infinite temperature.

From Eq.~\ref{eq:Q2}, imposing that $(a)=(b)$, we have that 
\begin{eqnarray}
 Q^2_\mathrm{self} &=& \frac{1}{V^2} \sum_{\vec x, \vec y}^V (\vec \sigma_{\vec x} \cdot \vec \sigma_{\vec y})^2 \\
                   &=& \frac{1}{V}  + \frac{1}{V^2} \sum_{\vec x \neq \vec y} (\vec \sigma_{\vec x} \cdot \vec \sigma_{\vec y})^2 \,.
\end{eqnarray}
The typical value of the scalar product $(\vec \sigma_{\vec x} \cdot \vec \sigma_{\vec y})^2$ at infinite 
temperature is trivially $1/m$, since it is equivalent to a random component of an $m$-dimensional unitary sphere.
Therefore the expectation value of the selfoverlap for a configuration at infinite temperature is
\begin{equation}
 E(Q^2_\mathrm{self}) = \frac{1}{V} + \frac{V-1}{mV} \sim \frac{1}{m} \,.
\end{equation}
This relation, that takes in account finite-size effects and 
was verified for random configurations, can be compared with
the selfoverlaps of the inherent structures in Table~\ref{tab:IS-m} (and with Fig.~\ref{fig:evol-qself}),
that are consistently different, since this observable has a non-trivial dependency on the temperature.

\end{document}